\documentclass{ws-ijmpb}

\usepackage{graphicx}
\usepackage{amsmath}
\usepackage{amsfonts}
\usepackage{amssymb}
\usepackage{mathptmx}      
\usepackage{latexsym}
\usepackage{hyperref}

\begin{document}

\markboth{Myong Chol Pak, Chol Won Ri, Hak Chol Pak, Kum Hyok Jong}
{Phonon Dispersion Relationship and Oxygen Isotope Effect in Superconductor LaFeAsO}

%
\catchline{}{}{}{}{}
%

\title{Phonon Dispersion Relationship and Oxygen Isotope Effect in Superconductor LaFeAsO}

\author{Myong Chol Pak, 
        Chol Won Ri\footnote{\email{cholwon\_ri@163.com}},  
    	Hak Chol Pak, 
    	Kum Hyok Jong}

\address{
           Department of Physics, Kim Il Sung University, Ryongnam Dong, Taesong District \\
           Pyongyang, Democratic People's Republic of Korea}
       
\maketitle

\begin{history}
	\received{13 November 2018}
\end{history}

\begin{abstract}
In this paper we calculate \textit{ab initially} the phonon dispersion relationship of the superconductor LaFeAsO and investigate a main property in the superconductor, the oxygen isotope effect. 
Based on this phonon dispersion relationship, we find the fact that an important reason of the oxygen isotope effect is connected with the phonon. 
This result agrees well with the experimental data where the power index of the oxygen isotope effect in the superconductor LaFeAsO is small. 
\end{abstract}

\keywords{Phonon dispersion relationship; Isotope effect; Superconductor}

\section{Introduction} \label{sec:intro}
High-temperature superconductors (HTSCs), which are the subjects of condensed matter physics, have been investigated for a long time, but the microscopic mechanisms have not been completely explained.
In particular, previous studies of the isotope effect in HTSC do  not give the clear solution about the reason why the power index of the isotope effect in these materials is small.\cite{Weyeneth_JSNM24,Alexandrov_NJP14,Mu_CPL25,Zhu_SST21,Shan_EPL83,Khasanov_PRB77}

In general, the isotope effect is a phenomenon where the critical temperature of superconductivity varies in inverse proportion to the square root of the isotope mass, and the existence of the isotope effect implies that the superconductivity is not related to pure Coulomb interaction and is related to the interaction between electron and lattice vibration.
In this reason, the interaction between electron and phonon is a very important object for resolving the mechanism of superconductivity.
Fr\"ohlich showed that the interaction between electron and phonon gave an attractive interaction between electrons that might be the reason of superconductivity.\cite{Frohlich_PR79}  
Fr\"ohlich's theory played a significant role for resolving the correct mechanism of superconductivity. 
Based on this result, in 1957, Bardeen, Cooper and Schrieffer (BCS) suggested the BCS theory, the successful microscope theory of superconductivity for the first time.\cite{Bardeen_PR108} 
The BCS theory pointed out the relationship between the critical temperature and the isotope mass is $\alpha  =  - d\ln T_c /d\ln M = 1/2$, which agrees well with the experiment result in simple metallic superconductors like Hg, Sn and Pb. 
Especially, the investigations for resolving the mechanism are intensified after the oxide high-temperature superconductor is discovered.\cite{Bednorz_ZPB64}
 
In the oxide high-temperature superconductor the fact that the carrier of superconductivity is the Cooper pair with electric charge $2e$ implies the reason of HTSC relates with the electron-phonon interaction and shows the existence of isotope effect like in the metallic superconductor. 
However, the power index of isotope effect $\alpha$ is not measured to 0.5, but to the very small value, 0.02, in the case for replacing $^{16} {\rm{O}}$ with $^{18} {\rm{O}}$ in HTSC of Y-series.\cite{Alexandrov_NJP14}  
This shows the reason of HTSC differs from one of the metallic superconductivity. 
Therefore, the non-phonon models are widely used for resolving the mechanism of HTSC.\cite{Hague_PRL98} 
However, some experiment results show the mechanism of HTSC is not independent of the electron-phonon interaction recently. 

The existence of c-axis polarized optical phonon in HTSC was suggested by using high resolution angle-resolved photoemission spectroscopy (ARPES).\cite{Bednorz_ZPB64}
This result provides a direct evidence for existing the electron-phonon interaction.

Also, the experiment results connected with optical scattering, neutron scattering, and tunnelling data show that the phonons play a significant role in HTSC.\cite{Lanzara_PRL96,Reznik_Nature440,Zhao_PRB75,Shim_PRL101,Zhao_PRL103,Kamihara_JACS30,Kamihara_ACS128}
Additionally, the analysis of the optical spectra of HTSC connected with multi-polaron absorption shows that the electron-phonon interaction is very important in those materials.\cite{Tempere_PRB64} 
However, these investigations have not been still resolved the reason of small power index of the isotope effect. 

In this paper we calculate ab initially the phonon dispersion relationship in the case of $^{16} {\rm{O}}$ and $^{18} {\rm{O}}$ for superconductor LaFeAsO respectively and investigated a main property in the superconductor, oxygen isotope effect.
On the basis of this phonon dispersion relationship we have found the fact that an important reason of the oxygen isotope effect is connected with the phonon.

This paper is organized as follows.
In Sec. \ref{sec:model_calculation}, we discuss the theoretical background for calculating the phonon dispersion relationship of the matter by using density functional perturbation theory (DFPT).
In Sec. \ref{sec:result_discussion}, the phonon dispersion relationship and the oxide isotope effect in the superconductor LaFeAsO are investigated by above mentioned method. 
In this section we calculate the longitudinal optical phone mode in the long wavelength limit and estimate the power index of the oxide isotope effect. 
In Sec. \ref{sec:summary}, summary and conclusion are presented.   

\section{Model and Calculation} \label{sec:model_calculation}

In general, the dynamical matrix should be considered in order to calculate the phonon dispersion relationship with DFPT.\cite{Pak_condmat1307,Lebegue_PRB75}

First, carrying out the normalization of the phonon displacement and considering the periodicity in the crystal, the following equation can be obtained; 
\begin{equation}
\label{eq:sc1}
u_{l\alpha i} = {1\over{\sqrt{M_\alpha}}}v_{l\alpha i},
\end{equation}
where $v_{l\alpha i}$ is $i^{th}$ component of the $\alpha^{th}$ atomic displacement in $l^{th}$ unit cell. 
Also, considering the translation symmetry in crystal, the result is
$$
u_{l\alpha i}  = u_{\alpha i} ({\bf{q}})e^{i{\bf{q}} \cdot {\bf{l}}},
$$
where a vector ${\bf{l}}$ indicates ${\bf{R}}_l$. 
Then above equation is as follows.
\begin{equation}
\label{eq:sc2}
u_{\alpha i} ({\bf{R}}_l ,{\bf{q}}) = u_{\alpha i} (0,{\bf{q}})e^{i{\bf{q}} \cdot {\bf{R}}_l }.
\end{equation}
In fact, the translation group is Abelian group and its irreducible representation is one-dimensional by Schur’s lemma.\cite{Schur_Book,Cornwell_Book} 
Applying some group operation, the result shows that the absolute value of this irreducible representation is one. 
Therefore, the equation is multiplied by the exponential-type factor, which is succeeded to Bloch's theorem for the wave function of the electron in the crystal.
By means of above discussion the dynamical matrix in ${\bf{q}}$ -space is 
\begin{equation}
\label{eq:sc3}
D_{ij}(\alpha\alpha',{\bf{q}}) = {1\over N}\sum\limits_{l,l'} {{1 \over{\sqrt{M_\alpha M_{\alpha'}}}}}\Phi_{ij}({\bf{l}}\alpha ,{\bf{l'}}\alpha')e^{-i{\bf{q}} \cdot ({\bf{l}}-{\bf{l'}})},
\end{equation}
where
\begin{equation}
\label{eq:sc4}
\Phi_{ij} ({\bf{l}}\alpha,{\bf{l'}}\alpha') = \Phi_{ij}^{el} ({\bf{l}}\alpha,{\bf{l'}}\alpha') + \Phi_{ij}^{ion} ({\bf{l}}\alpha ,{\bf{l'}}\alpha'),
\end{equation}
\begin{equation}
\label{eq:sc5}
\Phi_{ij}^{el}({\bf{l}}\alpha,{\bf{l'}}\alpha') = 2\sum\limits_{v,{\bf{k}}} {\left\langle {\psi_{v{\bf{k}}}} \right|{{\partial ^2 V_{ext} } \over {\partial u_{l\alpha i}^* \partial u_{l'\alpha'j}^{}}}} \left| {\psi _{v{\bf{k}}}} \right\rangle  + 2\sum\limits_{v,{\bf{k}}} {\left\langle {{{\partial \psi_{v{\bf{k}}}} \over {\partial u_{l\alpha i}^{}}}} \right|{{\partial V_{ext}} \over {\partial u_{l'\alpha'j}^{}}}} \left| {\psi_{v{\bf{k}}}} \right\rangle  + c.c..
\end{equation}
Then, $D_{ij}$ can be divided into the following three parts.
\begin{equation}
\label{eq:sc6}
D_{ij} (\alpha\alpha',{\bf{q}}) = D_{ij}^{(1)} (\alpha\alpha',{\bf{q}}) + D_{ij}^{(2)} (\alpha\alpha',{\bf{q}}) + D_{ij}^{(3)} (\alpha\alpha ',{\bf{q}}),
\end{equation}
where $D_{ij}^{(1)} (\alpha \alpha ',{\bf{q}})$  is a part connected with second-order derivatives of external potential and it is expressed by Eqs.~(\ref{eq:sc3}), (\ref{eq:sc5}) and $\sum\limits_{\bf{l}} {e^{i({\bf{q}} - {\bf{q'}}) \cdot {\bf{l}}} }  = N\delta ({\bf{q}} - {\bf{q'}})$ as follows.
\begin{equation}
\label{eq:sc7}
\begin{aligned}
D_{ij}^{(1)} (\alpha \alpha ',{\bf{q}}) =\ {2 \over {N\sqrt {M_\alpha  M_{\alpha '} } }}\sum\limits_{l,l'} {\sum\limits_{v,{\bf{k}}} {\left\langle {\psi _{v{\bf{k}}} } \right|{{\partial ^2 V_{ext} } \over {\partial u_{l\alpha i}^* \partial u_{l'\alpha 'j}^{} }}} \left| {\psi _{v{\bf{k}}} } \right\rangle e^{ - i{\bf{q}} \cdot ({\bf{l}} - {\bf{l'}})} }\\
 ={{2N} \over {\sqrt {M_\alpha  M_{\alpha '} } }}\sum\limits_{v,{\bf{k}}} {\left\langle {\psi _{v{\bf{k}}} } \right|{{\partial ^2 V_{ext} } \over {\partial u_{\alpha i}^* ({\bf{q}} = 0)\partial u_{l\alpha 'j}^{} ({\bf{q}} = 0)}}} \left| {\psi _{v{\bf{k}}} } \right\rangle. 
\end{aligned}
\end{equation}
Also, $D_{ij}^{(2)}$, a part connected with response of wave function due to the phonon perturbation is
\begin{equation}
\label{eq:sc8}
\begin{split}
D_{ij}^{(2)} (\alpha \alpha ',{\bf{q}})& =\ {2 \over {N\sqrt {M_\alpha  M_{\alpha '} } }}\sum\limits_{l,l'} {\sum\limits_{v,{\bf{k}}} {\left[ {\left\langle {{{\partial \psi _{v{\bf{k}}} } \over {\partial u_{l\alpha i}^{} }}} \right|{{\partial V_{ext} } \over {\partial u_{l'\alpha 'j}^{} }}\left| {\psi _{v{\bf{k}}} } \right\rangle  + c.c.} \right]} } e^{ - i{\bf{q}} \cdot ({\bf{l}} - {\bf{l'}})}\\ 
&={{2N} \over {\sqrt {M_\alpha  M_{\alpha '} } }}\sum\limits_{v,{\bf{k}}} {\left[ {\left\langle {{{\partial \psi _{v{\bf{k}}} } \over {\partial u_{l\alpha i}^{} }}} \right|{{\partial V_{ext} } \over {\partial u_{l'\alpha 'j}^{} }}\left| {\psi _{v{\bf{k}}} } \right\rangle  + c.c.} \right]}. 
\end{split}
\end{equation}
Finally, $D_{ij}^{(3)}$, a part connected with Coulomb interaction between ions is estimated by the Ewald sum method. 
This method by which Madelung's constant in the solid crystal is originally calculated is a very efficient method for investigating Coulomb interaction between ions. 
In this case, the electrostatic potential is divided as follows. 
$$
\varphi  = \varphi _1  + \varphi _2, 
$$
$$
\varphi _1  = \varphi _a  - \varphi _b, 
$$
$$
e\varphi _a  = {{2\pi Ne^2 } \over \Omega }\sum\limits_{{\bf{G}} \ne 0} {{{e^{ - G^2 /(4\eta )} } \over {G^2 }}} \left| {\sum\limits_\alpha  {Z_\alpha  e^{i{\bf{G}} \cdot {\bf{\tau }}_\alpha  } } } \right|^2,
$$
$$
e\varphi _b  = Ne^2 \sqrt {{\eta  \over \pi }} \sum\limits_\alpha  {Z_\alpha ^2 }, 
$$
$$
e\varphi _2  = {{Ne^2 } \over 2}\sum\limits_{\alpha ,\alpha '} {\sum\limits_{\bf{R}} {{{Z_\alpha  Z_{\alpha '} erfc(\sqrt \eta  \left| {{\bf{\tau }}_\alpha   - {\bf{\tau }}_{\alpha '}  - {\bf{R}}} \right|)} \over {\left| {{\bf{\tau }}_\alpha   - {\bf{\tau }}_{\alpha '}  - {\bf{R}}} \right|}}} }  - {{\pi e^2 } \over {\Omega \eta }}(\sum\limits_\alpha  {Z_\alpha  } )^2, 
$$
$$
\Phi _{ij}^{ion}  = e\varphi  = e(\varphi _a  - \varphi _b  + \varphi _2 ).
$$
Then, $D_{ij}^{(3)}$  is
\begin{equation}
\label{eq:sc9}
\begin{split}
D_{ij}^{(3)} (\alpha \alpha ',{\bf{q}}) &=\ {{4\pi e^2 } \over {\Omega \sqrt {M_\alpha  M_{\alpha '} } }}\sum\limits_{\scriptstyle {\bf{G}} \ne 0 \hfill \atop 
	\scriptstyle {\bf{q}} + {\bf{G}} \ne 0 \hfill}  {{{e^{ - ({\bf{q}} + {\bf{G}})^2 /(4\eta )} } \over {({\bf{q}} + {\bf{G}})^2 }}} Z_\alpha  Z_{\alpha '} e^{i({\bf{q}} + {\bf{G}}) \cdot ({\bf{\tau }}_\alpha   - {\bf{\tau }}_{\alpha '} )} (q_i  + G_i )(q_j  + G_j )\\
&- {{2\pi e^2 } \over {\Omega M_\alpha  }}\sum\limits_{{\bf{G}} \ne 0} {{{e^{ - G^2 /(4\eta )} } \over {G^2 }}} [Z_\alpha  \sum\limits_{\alpha ''} {Z_{\alpha ''} e^{i{\bf{G}} \cdot ({\bf{\tau }}_\alpha   - {\bf{\tau }}_{\alpha ''} )} G_i G_j  + c.c.} ]\delta _{\alpha \alpha '} \\
&+ {{e^2 } \over {\sqrt {M_\alpha  M_{\alpha '} } }}\sum\limits_{\bf{R}} {Z_\alpha  Z_{\alpha '} e^{i{\bf{q}} \cdot {\bf{R}}} [\delta _{\alpha \beta } f_2 (x) + f_1 (x)x_\alpha  x_\beta  ]}, 
\end{split}
\end{equation}
where
$$
{\bf{x}} \equiv {\bf{\tau }}_s  - {\bf{\tau }}_t  - {\bf{R}},
$$
\begin{equation}
\label{eq:sc10}
\begin{split}
&f_1 (x) = {{3erfc(\sqrt \eta  x) + 2\sqrt {{\eta  \over \pi }} x(3 + 2\eta x^2 )e^{ - \eta x^2 } } \over {x^5 }}.\\
&f_2 (x) = {{ - erfc(\sqrt \eta  x) - 2\sqrt {{\eta  \over \pi }} xe^{ - \eta x^2 } } \over {x^3 }}.
\end{split}
\end{equation}
From the dynamical matrix, the phonon dispersion relationship and the vibration mode can be obtained by using Eq.~(\ref{eq:sc6}).

In order to determine the phonon dispersion relationship in polar matter exactly, the contributions of longitudinal optical (LO) mode and transverse optical (TO) mode are considered in long wavelength limit.

In fact, the property in polar matter can be handled as one in a nonpolar matter in finite wavelength. 
In ${\bf{q}} \ne 0$ region, the moment of local dipoles in optical mode and acoustic mode vanish. 
However, the case of optical mode in long wavelength limit (${\bf{q}} \to 0$) is different from above one, because the centre of mass between the positive ion and the negative ion does not move.
In this case the moment of local dipoles in TO mode vanish, but in LO mode for asymmetry in the distribution of positive and negative ions. 
Therefore, some splitting between LO mode and TO mode is occurred in long wavelength limit. 
In polar matter, this is just called LO-TO splitting occurred in long wavelength limit. 
From this requirement, we should find the method for considering LO-TO splitting of long wavelength limit in polar matter. 

Meanwhile, the electric displacement in Gaussian unit is 
\begin{equation}
\label{eq:sc11}
D_\alpha   = E_\alpha   + 4\pi P_\alpha,
\end{equation}
where $P_\alpha$  is
\begin{equation}
\label{eq:sc12}
\begin{split}
P_\alpha  & =\left. {\sum\limits_{b,k',\beta } {{{\partial P_\alpha  } \over {\partial u_{bk'\beta } }}} } \right|_{\varepsilon  = 0} u_{bk'\beta }  + \left. {\sum\limits_\beta  {{{\partial P_\alpha  } \over {\partial \varepsilon _\beta  }}} } \right|_{u = 0} \left| E \right|n_\beta\\
& = {1 \over {\Omega _0 }}\sum\limits_{b,k',\beta } {Z_{bk'\beta }^* u_{bk'\beta }  + \sum\limits_\beta  {\chi _{\alpha \beta }^\infty  } \left| E \right|n_\beta}.
\end{split}
\end{equation}
Then, Eq.~(\ref{eq:sc11}) changes as follows.
\begin{equation}
\label{eq:sc13}
\begin{split}
D_\alpha   &= E_\alpha   + 4\pi P_\alpha   = E_\alpha   + {{4\pi } \over {\Omega _0 }}\sum\limits_{b,k',\beta } {Z_{bk'\beta }^* u_{bk'\beta }  + 4\pi \sum\limits_\beta  {\chi _{\alpha \beta }^\infty  } \left| E \right|n_\beta  }\\
&= {{4\pi } \over {\Omega _0 }}\sum\limits_{b,k',\beta } {Z_{bk'\beta }^* u_{bk'\beta }  + \left| E \right|\sum\limits_\beta  {(\delta _{\alpha \beta }  + 4\pi \chi _{\alpha \beta }^\infty  )} n_\beta  }. 
\end{split}
\end{equation}
Introducing the equation
\begin{equation}
\label{eq:sc14}
\varepsilon _{\alpha \beta }^\infty   = \delta _{\alpha \beta }  + 4\pi \chi _{\alpha \beta }^\infty,
\end{equation}
the electric displacement is
\begin{equation}
\label{eq:sc15}
D_\alpha   = {{4\pi } \over {\Omega _0 }}\sum\limits_{b,k',\beta } {Z_{bk'\beta }^* u_{bk'\beta }  + \left| E \right|\sum\limits_\beta  {\varepsilon _{\alpha \beta }^\infty  } n_\beta  }, 
\end{equation}
where $E_\alpha   = \left| E \right|n_\alpha$  and ${\bf{\hat n}} = (n_\alpha,n_\beta,n_\gamma)$ is the direction vector. 
From Maxwell's equation, $\sum\limits_\alpha  {n_\alpha  D_\alpha  }  = 0$. 
By Eq.~(\ref{eq:sc15}) the absolute value of the electric field is
\begin{equation}
\label{eq:sc16}
\left| E \right| =  - {{4\pi } \over {\Omega _0 }}{{\sum\limits_{b,k',\alpha ,\beta } {u_{bk'\beta } Z_{k'\beta \alpha }^* n_\alpha  } } \over {\sum\limits_{\alpha \beta } {n_\alpha  \varepsilon _{\alpha \beta }^\infty  } n_\beta  }}.
\end{equation}
Given the electric field in the long wavelength limit, the equation of motion considering Born effective charge is
\begin{equation}
\label{eq:sc17}
\begin{split}
F_{0k,\alpha }  &=\  - \sum\limits_{b,k',\beta } {C_{k\alpha k'\beta }^{TO} (0,b)u_{bk'\beta }  + \left| E \right|\sum\limits_\beta  {Z_{k'\beta \alpha }^* } n_\beta  }\\
& =  - \sum\limits_{b,k',\beta } {u_{bk'\beta } [C_{k\alpha k'\beta }^{TO} (0,b) + {{4\pi } \over {\Omega _0 }}{{\sum\limits_{\alpha '\beta '} {Z_{k\beta '\alpha }^* Z_{k'\beta \alpha '}^* n_{\alpha '} n_{\beta '} } } \over {\sum\limits_{\alpha '\beta '} {n_{\alpha '} \varepsilon _{\alpha '\beta '}^\infty  } n_{\beta '} }}} ]\\
& =  - \sum\limits_{b,k',\beta } {u_{bk'\beta } [C_{k\alpha k'\beta }^{TO} (0,b) + {{4\pi } \over {\Omega _0 }}{{(\sum\limits_{\alpha '} {Z_{k'\beta \alpha '}^* n_{\alpha '} )(\sum\limits_{\beta '} {Z_{k\beta '\alpha }^* n_{\beta '} } )} } \over {\sum\limits_{\alpha '\beta '} {n_{\alpha '} \varepsilon _{\alpha '\beta '}^\infty  } n_{\beta '} }}} ],
\end{split}
\end{equation}
where second term is connected with LO-TO splitting.

On the whole, the dynamical matrix including LO-TO splitting in polar matter is
\begin{equation}
\label{eq:sc18}
D_{ij}^{} (\alpha \alpha ',{\bf{q}}) = D_{ij}^{an} (\alpha \alpha ',{\bf{q}}) + D_{ij}^{na} (\alpha \alpha ',{\bf{q}} \to 0),
\end{equation}
where the non-analytic part connected with LO-TO splitting is
\begin{equation}
\label{eq:sc19}
D_{ij}^{na} (\alpha \alpha ',{\bf{q}} \to 0) = {{4\pi e^2 } \over {\Omega M_\alpha  M_{\alpha '} }}{{(\sum\limits_l {Z_{i\alpha l}^* n_l )(\sum\limits_m {Z_{j\alpha 'm}^* n_m } )} } \over {\sum\limits_{lm} {n_l \varepsilon _{lm}^\infty  } n_m }}.
\end{equation}
When the dynamic matrix is determined, the phonon dispersion relation $\omega=\omega(\mathbf{q})$ can be obtained by solving the following eigenvalue problem.
\begin{equation}
\label{eq:sc20}
\det \left\| {{D}_{ij}}(\alpha {\alpha }',\mathbf{q})\text{-}{{\omega }^{2}}(\mathbf{q}){{\delta }_{ij}} \right\|=0.
\end{equation}

\section{Result and Discussion}
\label{sec:result_discussion}

We apply the above method for calculating the phonon dispersion relationship in the superconductor LaFeAsO. 
This substance crystallizes in a tetragonal crystal structure and its crystal structure is layered with La-O and As-As layers.\cite{Singh_PRL100} 
The lattice constants in LaFeAsO are $a = 0.4035{\rm{nm}}$ and $c = 0.8741{\rm{nm}}$, and two As-Fe-As angles are $120.2^\circ$ and $104.4^\circ$. 
Also, the distance between Fe and As is 0.2327 nm, and the distance between Fe and Fe is 0.2854nm. 
The Fe-Fe distance is short enough, so that direct Fe-Fe hopping cannot be cancelled. 
Additionally, the distance between As and As is 0.3677nm, across the Fe layer.
Because this substance has the special crystal structure as above mentioned, the world-wide interests in it are getting higher. 

We carry out the calculation by using local spin density approximation (LSDA) in Quantum Espresso-4.04 code complied on Linux OS. 
This phonon dispersion relationship is obtained by the cutting energy 50Ry with above described linear response method.

\begin{figure}[!ht]
	\begin{center}
		\includegraphics[clip=true,scale=0.7]{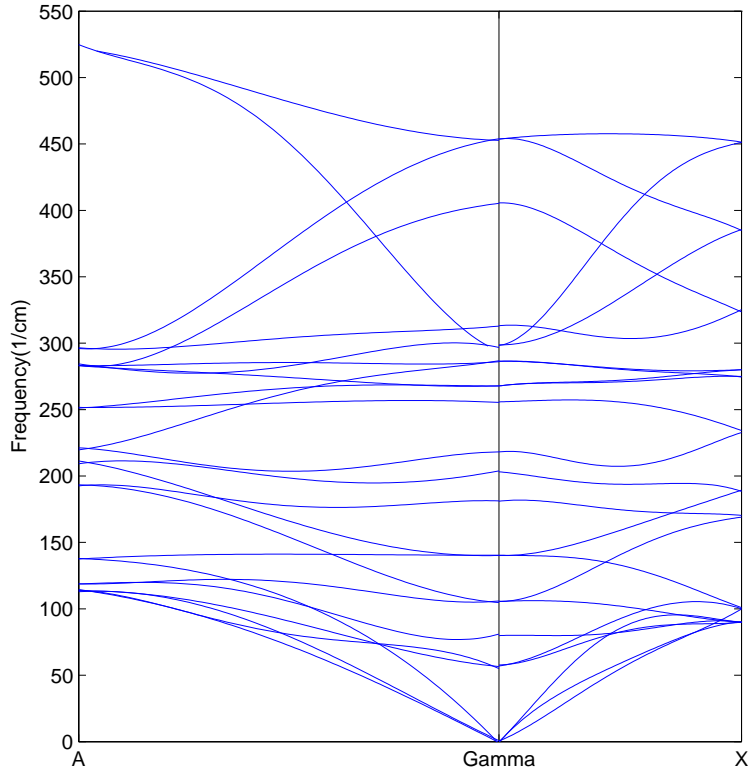}		
		\caption{\label{fig:ox-16}(Colour online) The phonon dispersion relationship in superconductor LaFeAsO including $^{16} {\rm{O}}$ in the vicinity of
			 $\Gamma$-point.}
	\end{center}
\end{figure}

\begin{figure}[!ht]
	\begin{center}
		\includegraphics[clip=true,scale=0.7]{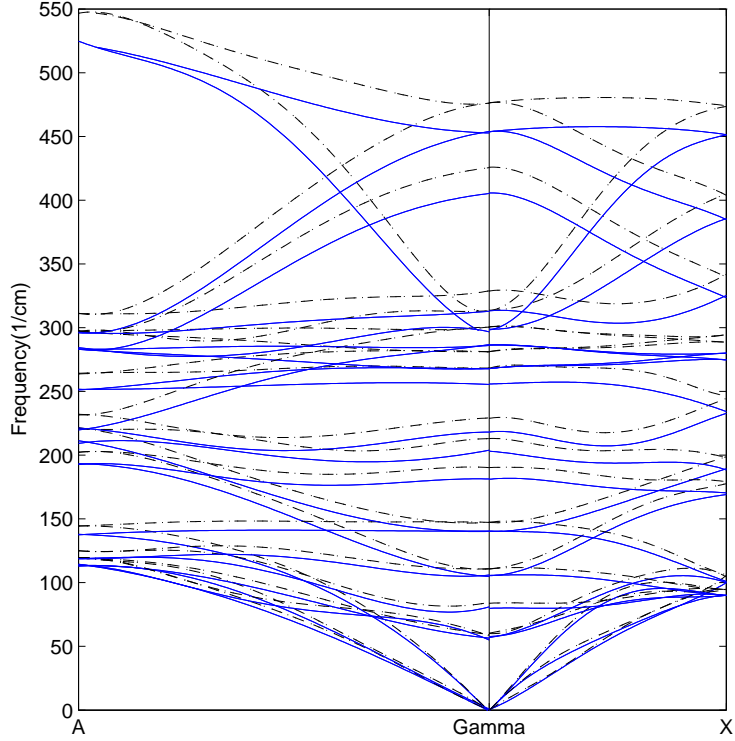}
		\caption{\label{fig:ox-18}(Colour online) The phonon dispersion relationships in LaFeAsO including $^{16} {\rm{O}}$ and $^{18} {\rm{O}}$. Solid line represents the case of $^{16} {\rm{O}}$ and dashed line represents the case of $^{18} {\rm{O}}$.}
	\end{center}
\end{figure}

In this paper, the phonon dispersion relationships of the superconductors including $^{16} {\rm{O}}$ and $^{18} {\rm{O}}$  are discussed to consider its oxide isotope effect, respectively (see Fig.~\ref{fig:ox-16} and \ref{fig:ox-18}). 
A unit cell has 8 atoms, so that 24 vibration modes appear. 
They consist of 3 acoustic modes, 7 LO modes and 14  TO modes.
The centre of our attention is LO mode in long wavelength limit. 
Therefore, the range in phonon dispersion relation is confined to A-point and X-point near $\Gamma$-point. 
As shown in Fig. \ref{fig:ox-16}, in the case including $^{16} {\rm{O}}$  the minimum frequency of LO mode in $\Gamma$-point is 56.4366${\rm{cm}}^{ - 1}$ and its maximum frequency is 453.5448${\rm{cm}}^{ - 1}$.


In Fig.~\ref{fig:ox-18}, we discuss the isotope effect with the way for calculating the case of $^{18} {\rm{O}}$ and comparing the case of $^{16} {\rm{O}}$ with the case of $^{18} {\rm{O}}$ . 
Really, the difference of the phonon dispersion in low frequency region is very small, but its difference in high frequency region is large. 
In the case of $^{18} {\rm{O}}$, the minimum frequency is 57.7884${\rm{cm}}^{ - 1}$, which differs by 1.3518${\rm{cm}}^{ - 1}$ from the case of $^{16} {\rm{O}}$. 

On the other hand, the maximum frequency is 473.2372${\rm{cm}}^{ - 1}$, which differs by 19.6924${\rm{cm}}^{ - 1}$ from the case of $^{16} {\rm{O}}$. 
However, the critical temperature of superconductor is connected with an inverse number of the frequency, so that only frequencies in low frequency region have a great influence on the critical temperature. 
In this reason, the oxide isotope effect exerts weakly in the superconductor LaFeAsO. 
This fact shows the experimental result~\cite{Alexandrov_NJP14} where the power index of the oxide isotope effect is small.
Therefore, we demonstrate that the oxide isotope effect in the superconductor is connected with the phonon.

It should be noted that not only phonon affects the superconductivity in LaFeAsO superconductor. 
In Ref. \refcite{Boeri_PRL101}, they already calculated the Tc by performing Eliashberg functions and estimating electron-phonon coupling constant in the superconductor LaFeAsO. 
According to the result in this paper, the calculated electron-phonon coupling constant is 0.21, which is $5 \sim 6$ times smaller than the coupling constant estimated for the experimental critical temperature. 
This result shows that the electron-phonon coupling is not sufficient to analyse the critical temperature in superconductor LaFeAsO. 
It also implies that the electron correlation of d-orbital electrons in Fe has an effect on the critical temperature. 
This can be resolved by considering the exchange-correlation term in density functional theory. 
Certain results related to it are suggested.\cite{Kaloni_EPL104,Zhang_PRL98} 
Although we do not necessarily imply that the oxygen isotope effect in this superconductor is only related to the phonon, the influence of the phonon on the oxygen isotope effect never can be ignored. 
The relation between the oxygen isotope effect and the electron correlation is a prospective task to be researched by using different methods, such as the method of upgrading exchange-correlation potential.

\section{Summary}
\label{sec:summary}
We have investigated the phonon dispersion relationship and the phonon vibration spectrum in the superconductor LaFeAsO including $^{16} {\rm{O}}$ and $^{18} {\rm{O}}$ by using ab initio method. 
Estimating LO mode in long wavelength limit, we have found that the power index of the oxide isotope effect is small. 
Therefore, a main mechanism of isotope effect connected with the phonon is demonstrated. 
This result implies the possibility for resolving the mechanism of HTSC with bipolaron.

\section*{Acknowledgements}
It is pleasure to thank Nam Hyok Kim, Chol Jun Yu, and Yong Hae Ko for useful discussions. 
This work is supported by the National Program on Key Science Research of Democratic People's Republic of Korea (Grant No. 18-1-3).

\end{document}